\documentclass[conference]{IEEEtran}
\usepackage{amssymb,amsmath,amsthm}
\usepackage[lined,boxed,commentsnumbered,ruled]{algorithm2e}
\usepackage{algorithmic} 
\usepackage{graphicx}
\usepackage{stmaryrd}
\SetSymbolFont{stmry}{bold}{U}{stmry}{m}{n}
\usepackage{multirow,url}
\usepackage{color,cite}
\usepackage{cite}
\usepackage{tikz,pgf}
\usetikzlibrary{fadings,shapes,arrows,shadows}
\usepackage{makecell}
\usepackage{array}
\usepackage{overpic} 
\usepackage{subfigure}
\usepackage{url}

\newtheorem{lemma}{Lemma}

\newtheorem{definition}{Definition}
\newtheorem{theorem}{Theorem}

\theoremstyle{plain}

\ifodd 1
\newcommand{\revv}[1]{{#1}} 
\newcommand{\com}[1]{\textbf{\color{red} (COMMENT: #1)}} 
\newcommand{\comg}[1]{\textbf{\color{green} (COMMENT: #1)}} 
\newcommand{\response}[1]{\textbf{\color{green} (RESPONSE: #1)}} 
\else

\newcommand{\revv}[1]{#1}

\newcommand{\com}[1]{}
\newcommand{\comg}[1]{}
\newcommand{\response}[1]{}
\fi

\def\x{\boldsymbol{x}}
\def\y{\boldsymbol{y}}
\def\z{\boldsymbol{z}}
\def\p{\boldsymbol{p}}
\def\b{\boldsymbol{b}}
\def\c{\boldsymbol{c}}

\def\pv{\boldsymbol{r}}
\def\t{\boldsymbol{t}}

\def\bu{\b_{\textsc{User}}}
\def\bt{\b_{\textsc{Task}}}

\def\I{\mathcal{I}}
\def\J{\mathcal{J}}

\def\D{\mathcal{K}}

\def\S{\mathcal{S}}

\def\eq{\triangleq}

\def\mech{\Omega}

\IEEEoverridecommandlockouts

\begin{document}

\title{A Double Auction Mechanism for Mobile Crowd Sensing with Data Reuse \vspace{-3mm}
}

\author{Xiaoru Zhang, Lin Gao, Bin Cao, Zhang Li, and Mengjing Wang
\thanks{Authors are with the School of Electronic and Information Engineering, Harbin Institute of Technology, Shenzhen, China. 
Lin Gao is the Corresponding Author. Email: gaol@hit.edu.cn.
}
\thanks{This work is supported by the National Natural Science Foundation of China  (Grant No. 61771162 and 61501211) and the Basic Research Project of Shenzhen (Grant No. JCYJ20160531192013063 and JCYJ20170307151148585).}
}

\maketitle

\begin{abstract}
Mobile Crowd Sensing (MCS) is a new paradigm of sensing, which can achieve a flexible and scalable sensing coverage with a low deployment cost, by employing mobile users/devices to perform sensing tasks.
In this work, we propose a novel MCS framework with data reuse, where multiple tasks with common data requirement can share (\emph{reuse}) the common data with each other through an MCS platform.
We study the optimal assignment of mobile users and tasks (with data reuse) systematically, under both information symmetry and asymmetry, depending on whether the user cost and the task valuation are public information.
In the former case, we formulate the assignment problem as a generalized Knapsack problem and solve the problem by using classic algorithms.
In the latter case, we propose a truthful and optimal double auction mechanism, built upon the above Knapsack assignment problem, to elicit the private information of both users and tasks and meanwhile achieve the same optimal assignment as under information symmetry.
Simulation results show by allowing data reuse among tasks, the social welfare can be increased up to $100\%\sim380\%$, comparing with those without data reuse.
We further show that the proposed double auction is \emph{not} budget balance for the auctioneer, mainly due to the data reuse among tasks.
To this end, we further introduce a reserve price into the double auction (for each data item) to achieve a desired tradeoff between the budget balance and the social efficiency.
\end{abstract}

\IEEEpeerreviewmaketitle


\section{Introduction}
\label{sec:introduction}


\subsection{Background and Motivations}

The proliferation of mobile devices (e.g., smartphones) with rich embedded sensors has led to a novel sensing paradigm known as \emph{Mobile Crowd Sensing (MCS)} \cite{MCSSurvey}, where mobile users/devices are employed to perform different sensing tasks.
By crowdsourcing the sensing capabilities of massive powerful mobile devices, this new sensing paradigm can achieve a high sensing coverage with a low deploying cost, hence has attracted a wide range of applications in environment, infrastructure, and community  monitoring (e.g., \cite{App-OpenSense, App-Atmos1, App-WeatherLah, App-OpenSignal, App-NoiseTube, App-Sensorly, App-Waze, App-Millennium, App-CarTel,App-SpotSwitch, App-ParkNet}).~~~~~~~~~

A typical MCS framework  mainly consists of the following three parts \cite{MCSSurvey}:
(i) a set of \emph{task planners}, who initiate sensing tasks with specific data requirements,
(ii) a set of \emph{mobile users}, who report their capabilities and interests for performing different tasks,
and (iii) an \emph{MCS platform}, who collects the information of tasks and users, and assigns tasks to users carefully.
When a user is assigned with a task, he performs the task accordingly using his device resource (e.g., CPU cycles and energy for sensing and processing data, bandwidth and energy for sending data to the task planner), which will incur a certain cost on the user.
Meanwhile, the user  can obtain a certain reward from the task planner via the platform according to a certain pre-defined payment rule.

Some prior works (e.g., \cite{SmartphoneCollaboration,DYang,TieLuo,YanminZhu,Lin,Jiang,Jiang2}) have studied the general MCS model with multiple tasks and multiple users from different aspects,
such as how to assign tasks to users efficiently, how to determine the rewards for users and the payments from tasks properly, and so on.
Most of the existing works (i.e., \cite{SmartphoneCollaboration,DYang,TieLuo,YanminZhu,Lin,Jiang}) focused on the MCS model \emph{without} data reuse among tasks, where the same data required by multiple tasks cannot be shared (reused) among tasks and has to be sensed distinctly for each task.
In practice, however, it is highly likely that different tasks require (hence can reuse) some common data \cite{Jiang2}.
For example, the weather data at a particular time and location may be required by the weather app (task 1), the travel app (task 2), and the road navigation app (task 3).
Thus, without data reuse, it is likely to cause duplicated data sensing and processing, leading to resource waste and performance degradation.
For this purpose, some practical MCS platforms such as PRISM~\cite{Platform1} and Medusa~\cite{Platform2} have allowed task planners to define data requirements in a standard language, such that the common data (requirement) of different tasks can be identified and reused potentially.

%
%

In \cite{Jiang2}, Jiang \emph{et al.} studied the MCS model with data reuse among tasks, which allows multiple tasks with the common data requirement to share the common data with each other.
They proposed a randomized auction mechanism, which is truthful in expectation and approximately optimal in term of social welfare.
In this work, we consider a similar MCS model which allows data reuse among tasks,
but our purpose is to design a mechanism which is \emph{strictly} truthful and optimal.

\begin{figure}[t]
\vspace{-4mm}
\centering
\includegraphics[width=0.47\textwidth]{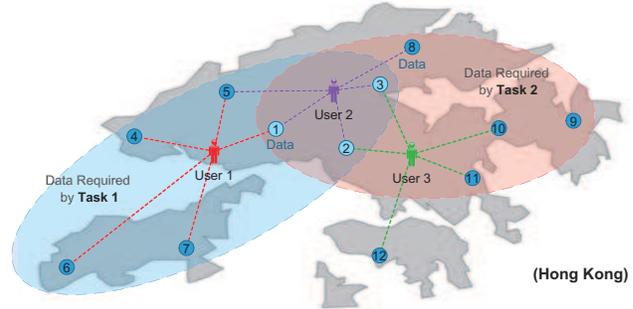}
\vspace{-4mm}
\caption{An MCS model with common data requirement among tasks.
Tasks $1$ and $2$ require the common data set $\{1,2,3\}$.}
\label{Fig:SystemModel}
\vspace{-4mm}
\end{figure}


\subsection{Model and Problem}

In this work, we study a general  {multi-task multi-user} MCS model with data reuse, where different tasks can have common data requirements and reuse the common data.
Specifically, each task is associated with a set of data items that it requires, and the overlap of different tasks' data sets is the common data requirement of those tasks.
Each user is associated with a set of data items that he can sense, and such a sensing capability depends on factors such as location, device capability, budget constraint, and so on.

Figure~\ref{Fig:SystemModel} illustrates such an MCS model with common data requirement (and data reuse) among tasks.
Each blue circle denotes a particular data item (e.g., specific information at a particular time and location), and the dash line between a user and a data item denotes that the user can sense the data item.
Each ellipse denotes the data items interested and required by each task.
Obviously, tasks $1$ and $2$ require the common data set $\{1,2,3\}$, hence can potentially reuse those data items $\{1,2,3\}$.
It is easy to see that our model generalizes the traditional model without data reuse in \cite{SmartphoneCollaboration,DYang,TieLuo,YanminZhu,Lin,Jiang},
as it can degenerate to the traditional model by simply viewing each common data required by multiple tasks as multiple virtual data, each associated with a particular task.

In such an MCS model, we are interested in the following \emph{Task-Data-User Assignment (TDU-A)} problem:
\begin{itemize}
  \item \emph{How to assign different users to sense different data of different tasks, aiming at maximizing the social welfare, taking the   data reuse among tasks into consideration?}
\end{itemize}
The \emph{social welfare} is defined as the difference between the  {total valuation} of all completed tasks and the  {total sensing cost} of all employed users.
More specifically, a task is completed if and only if all of its required data items have been sensed successfully, and the completion of a task will generate a certain \emph{valuation} for the task planner.
A user will incur certain sensing cost when he is scheduled to sense a data item, and each user is associated with a \emph{budget}, denoting the maximum sensing cost that the user can afford.

Solving the above problem is challenging due to the following reasons.
First, from the technical perspective, a simplified version of the problem (with a single user) is a Knapsack problem \cite{knapsack} (which is NP-hard), where the total user budget can be viewed as the knapsack capacity, and the sensing cost for each data item can be viewed as the weight of the item.
Besides, involving the intermediate layer of data (between tasks and users) will further complicate the problem.
Second, from the economic perspective, users may not be willing to report their sensing costs and budgets truthfully, and task planners may not be willing to report their valuations truthfully.
Hence, a well-designed incentive mechanism (e.g., VCG auction \cite{vcg}) is necessary for eliciting the private information of both users and task planners and making the assignment.

\subsection{Solution and Contributions}

We will solve the problem under both information \emph{symmetry} and \emph{asymmetry}, depending on whether the user sensing cost and budget and the task valuations are public information.
In the former case, all information are public and can be observed by the platform.
We formulate the assignment problem as a generalized Knapsack problem and solve the problem by using classic algorithms.
In the latter case, the sensing cost and budget are private information of each user, and the task valuation is private information of each task planner, both of which cannot be observed by the platform.
We propose a truthful \emph{double auction} with the platform as the \emph{auctioneer} and users and task planners as \emph{bidders},
which can elicit the private information of users and task planners creditably, and meanwhile achieve the same optimal social welfare as under information symmetry.

We further show that the proposed double auction is \emph{not} budget balance for the auctioneer (platform),
in the sense that the total payment from all task planners may be smaller than the total reward to all employed users.
This implies that the platform may need additional budget for organizing such an auction, which is not desirable in practice.
To avoid this, we further introduce a \emph{reserve price} to restrict the minimum payment (of tasks) for each data item.
We show that a desired tradeoff between the budget balance and the social efficiency can be achieved by turning the reserve price finely.

In summary, the main results and  key contributions of this work are summarized as follows.

\begin{itemize}
\item \emph{Novel MCS Model:}
We study a novel MCS model which allows multiple tasks to reuse the common data.
Comparing with the existing models without data reuse, this new model can reduce the duplicated data sensing and processing, hence increase the social efficiency.

\item \emph{Truthful and Optimal Auction Design:}
We study the optimal assignment problems under both information symmetry and asymmetry.
In particular, we propose a truthful and optimal double auction under information asymmetry, which can elicit the private information of users/tasks  creditably and achieve the same optimal social welfare as under information symmetry.

\item \emph{Performance Evaluations:}
Simulation results show by allowing data reuse among tasks, the social welfare can be increased up to $100\%\sim380\%$, comparing with those without data reuse.
We further compare our approach with that in \cite{Jiang2}, and show that our proposed double auction (which is provably optimal) has an average of $10\%$ performance gain over the randomized auction in  \cite{Jiang2} (which is approximately optimal).

\item \emph{Observations and Insights:}
We show that allowing data reuse among tasks will reduce the competition of tasks, hence reduce the payments of task planners.
This may lead to the undesired outcome of budget unbalance for the platform.
We further show that a well-designed reserve price can achieve a desired tradeoff between the budget balance and the social efficiency.

\end{itemize}

The rest of the paper is organized as follows.
In Section \ref{sec:model}, we present the system model.
In Section \ref{sec:symmetry} and \ref{sec:asymmetry}, we analyze the problem under information symmetry and asymmetry, respectively.
We present the simulation results in Section \ref{sec:simulation}, and finally conclude in Section \ref{sec:conclusion}.


\section{System Model}
\label{sec:model}

\subsection{Network Model}

We consider a multi-task multi-user MCS model, consisting of
a set $\I = \{1, \cdots,I\}$ of mobile users,  a set $\J=\{1, \cdots,J\}$ of tasks, and a set $\D=\{1, \cdots,K\}$ of target data items.
Each data item corresponds to a specific information at a particular  location and time.\footnote{For example, a data item can be the temperature of a particular location at 11:00pm everyday, the traffic of a highway at a particular time, or a raw sensor reading such as GPS, light sensor, accelerometer, and gyroscope.}
Each task $j \in \J $ is associated with a \emph{data requirement}, i.e., a set $\D_j \subseteq \D $ of data items that it requires.
Each user $i \in \I$ is associated with a \emph{sensing capability}, i.e., a set $\S_i \subseteq \D $ of data items that he can sense.
Note that different tasks may have  common data requirement, i.e., $\D_{j_1} \bigcap \D_{j_2} \neq \emptyset$, and can reuse the common data potentially.

In the example of Figure~\ref{Fig:SystemModel}, we have: $I=3$ users,  $J=2$ tasks, and $K = 12$ target data items. The data requirements of both tasks are $\D_1 = \{1 \mbox{--} 7\}$ and $\D_2 = \{1 \mbox{--} 3, 8 \mbox{--} 11 \}$. The sensing capabilities of three users are $\S_1 = \{1,4 \mbox{--} 7\}$, $\S_2 = \{1 \mbox{--} 3, 5, 8\}$, and $\S_3 = \{2,3,10 \mbox{--} 12\}$.
Obviously,  tasks $1$ and $2$ can reuse the common data set $ \D_1 \bigcap \D_2 = \{1,2,3\}$.

The system operates in the following way.
First, each task planner publishes the sensing task on the platform, indicating the data requirement and   task valuation (under certain incentive mechanism). Each user reports the sensing interest on the platform, indicating the
sensing capability, sensing cost, and sensing budget (under certain incentive mechanism).
Then, based on the information reported by task planners and users, the platform decides the task-data-user assignment, aiming at maximizing the social welfare.

\subsection{User Model}

Each user $i \in \I $ is associated with (i) a sensing capability $\S_i \subseteq \D $, denoting the set of data items that he can sense,
and (ii) a sensing cost vector $\c_i \eq (c_{i,k}, k\in \S_i )$, denoting the cost for sensing each data item in  $\S_i$.\footnote{For a data item $k $ not in a user $i$'s sensing capability $\S_i$, we can say that his sensing cost for
data item $k$ is infinite, i.e., $c_{i,k} = \infty, \forall k \notin \S_i$.}
Such a sensing cost mainly consists of the resource consumption for moving to the target location, collecting and processing the data, and transmitting the data to the task planner.
Note that a user may have different sensing costs for different data items, due to, for example, the different distances to those data items or the different capabilities for processing different data.
Moreover, each user $i$ is associated with a sensing \emph{budget} $C_i > 0$, capturing  the maximum resource that can be used for sensing.
Such a budget may depend on factors such as the user's own service requirement and resource availability.
{For example, a user with a heavy burden of service or a low available device resource may assign a low budget for sensing.}

Let $x_{i,k} \in \{0, 1\}$ denote whether a user $i \in \I$ is scheduled to sense a data item $k \in \S_i$, and $\x_i \eq (x_{i,k}, k\in \S_i)$ denote the sensing scheduling vector of user $i$.
Then, we have the following budget constraint for user $i$:
\begin{equation} \label{eq:budget}
 \sum_{ k \subseteq \S_i  }  x_{i,k} \cdot c_{i,k} \leq C_i
\end{equation}

Given the feasible scheduling vectors of all users, i.e., $\x \eq (\x_i, i\in\I)$, the total incurred sensing cost on all users is:
\begin{equation}\label{eq:Cx}
  C(\x) = \sum_{i\in \I} \sum_{ k \subseteq \S_i  }  x_{i,k} \cdot c_{i,k} .
\end{equation}

\subsection{Data Model}

A data item is sensed successfully, if and only if it is sensed by at least one user.
Let $\I_k \eq \{ i\in\I\ |\ k\in\S_i \}$ denote the set of users that can sense a data item $k$, and $y_k \in \{0, 1\}$ denote whether a data item $k \in \D$ is sensed successfully.
Then,
\begin{equation}\label{eq:ykkk}
 y_k = \max \{ x_{i,k} ,\ \forall i \in \I_k \}.
\end{equation}
In the example of Figure \ref{Fig:SystemModel}, we have:
(i) $\I_1 = \{1,2\}$ and $y_1 = \max\{ x_{1,1}, x_{2,1} \}$ for data item $1$,
(ii) $\I_2 = \{2,3\}$ and $y_2 = \max\{ x_{2,2}, x_{3,2} \}$ for data item $2$, and
(iii) $\I_4 = \{1\}$ and $y_4 = x_{1,4}$ for data item $4$.
For convenience, we denote $\y \eq (y_k, k \in \D) $ as the sensing indicators of all data items.

\subsection{Task Model}

Each task $j \in \J $ is associated with (i) a data requirement $\D_j \subseteq \D $, denoting the set of data items that it requires,
and (ii) a task valuation $v_j > 0$, denoting the value of task for the task planner when the task is completed.
A task $j$ is \emph{completed} if and only if all of its required data items in $\D_j $ have been successfully sensed by at least one user.
Let $z_j \in \{0, 1\}$ denote whether a task $j \in \J$ is completed.
Then, for each task $j\in\J$, we have the following completion indicator:
\begin{equation}\label{eq:zj}
z_j  =  \min\{ y_k,\  \forall k \in \D_j\}.
\end{equation}
In the example of Figure \ref{Fig:SystemModel}, we have:
(i) $\D_1 = \{1 \mbox{--} 7\}$ and $z_1 = \min\{ y_1  \cdots  y_7\}$   for task $1$
and (ii) $\D_2 = \{1 \mbox{--} 3, 8 \mbox{--} 11 \}$ and
$z_2 = \min\{ y_1, y_2, y_3,y_8 \cdots y_{11}\}$ for task $2$.

Given the task completion indicators of all tasks, i.e., $\z \eq (z_j,j\in\J)$, the total achieved valuation is:
\begin{equation}\label{eq:Vz}
  V(\z) = \sum_{j\in \J} z_j \cdot v_j .
\end{equation}






\subsection{Problem Formulation}

The social welfare $W(\boldsymbol{x}, \boldsymbol{z})$ is defined as the difference between the
total   valuation $V(\boldsymbol{z})$ of all completed  tasks and the total
 sensing cost $C(\boldsymbol{x})$ of all employed   users, i.e.,
\begin{equation}\label{eq:Sxz}
W(\boldsymbol{x}, \boldsymbol{z}) = V(\boldsymbol{z}) - C(\boldsymbol{x}).
\end{equation}

Our purpose is to decide the best task-data-user assignment  $\{\x, \y, \z\}$  that maximizes the social welfare, taking the potential data reuse among tasks and the budget constraints of users into considerations.
Specifically, we can formulate the joint task-data-user assignment problem (A1) as follows.
\begin{align}\label{eq:p1}
\mbox{A1:} & & \max  & ~~~ V(\boldsymbol{z}) - C(\boldsymbol{x})
\notag\\
& & \mbox{s.t.} & ~~~  \eqref{eq:budget} \eqref{eq:ykkk} \eqref{eq:zj}, \quad \forall i \in \I, j \in \J, k \in \D;
\notag\\
& & \mbox{var.} & ~~~  x_{i,k} \in \{0,1\}, \quad \forall i \in \I , k\in\S_i;
\notag\\
& & & ~~~   z_j \in \{0, 1\},\quad \forall j \in \J ;\notag\\
& &  & ~~~ y_k \in \{0, 1\},\quad \forall   k \in \D.\notag
\end{align}
Here $\boldsymbol{y} =  (y_k, k\in\D)$ is an intermediate variable indicating  whether a data item is sensed successfully, which connects the tasks and users.
It is easy to see that Problem P1 is a binary integer linear programming problem.


However, solving the Problem A1 is challenging due to the following reasons.
First, it is a generalized Knapsack problem, as a simplified version of the problem (with a single user) is a Knapsack problem  \cite{knapsack} (which is NP-hard), where the total user budget can be viewed as the knapsack capacity, and the sensing cost for each data item can be viewed as the weight of the item.
Second, it involves an intermediate data layer in the assignment of users and tasks, leading to a \emph{three-layer} model, which makes the problem even more complicated and challenging.
Third, it requires the complete information of the whole system, including the data requirements and valuations of all tasks as well as the sensing capabilities, sensing costs, and budgets of all users.
However, such information can be private information in practice, and task planners or users may not be willing to report their private information truthfully.
Hence, we need to design a proper incentive mechanism to elicit such private information.

In what follows, we first solve   Problem A1 under information \emph{symmetry} where all information is public information (Section \ref{sec:symmetry}).
Then we study   Problem A1 under information \emph{asymmetry} where the above mentioned information is private information of task planners and users (Section \ref{sec:asymmetry}).

\section{Information Symmetry}
\label{sec:symmetry}

In this section, we consider the information \emph{symmetry} scenario, where all information is public information and can be observed by the platform.
Hence, the key problem is to solve the optimal task-data-user assignment by the Problem A1.

We first show that the Problem A1 is a Knapsack problem \cite{knapsack}.
To show this, we consider a simplified model with $J$ tasks, each requiring a distinct data item, and $I=1$ user who can sense all data items.
Then the problem becomes the following: \emph{select the tasks (or data items) to be completed within the user budget}.
Let us view the user budget as the knapsack \emph{capacity}, the sensing cost for each data item as the \emph{weight} of the item, and the valuation of each data item (task) as the \emph{value} of the item.
It is easy to see that the problem in this simplified model is exactly a Knapsack problem.
Thus, the general case of Problem A1 is a generalized Knapsack problem.
Note that there are many efficient algorithms for solving Knapsack problems, either optimally or sub-optimally \cite{knapsack}.
Due to space limit, we will not go into the details of these algorithms in this work.

We further notice that it is still challenging to apply the classic algorithms to solve the Problem A1, mainly due to the $\min$ and $ \max$ operations in the equality constraints \eqref{eq:ykkk} and \eqref{eq:zj}.
Hence, it is necessary to transform these constraints into other equivalent forms. Formally,

\begin{lemma}\label{lemma:1}
The equality constraint in \eqref{eq:ykkk} is equivalent to the following constraints:
\begin{equation}\label{eq:ykkk-eq}
 y_k  \geq  x_{i,k},\ \forall i \in \I_k,
 \mbox{~~~~and~~~~}
 y_k  \leq \sum_{i \in \I_k} x_{i,k} .
\end{equation}
\end{lemma}


\begin{lemma}\label{lemma:2}
The constraint in \eqref{eq:zj} can be relaxed to the following constraints without affecting the optimal solution:
\begin{equation}\label{eq:zj-eq}
 z_j \leq  y_k,\ \forall k \in \D_j.
\end{equation}
\end{lemma}


Based on the above lemmas, we can transform the Problem A1 into the following equivalent Problem A2.
\begin{align}\label{eq:p1-eq}
\mbox{A2:} & & \max  & ~~~ V(\boldsymbol{z}) - C(\boldsymbol{x})
\notag\\
& & \mbox{s.t.} & ~~~  \eqref{eq:budget} \eqref{eq:ykkk-eq} \eqref{eq:zj-eq}, \quad \forall i \in \I, j \in \J, k \in \D;
\notag\\
& & \mbox{var.} & ~~~  x_{i,k} \in \{0,1\}, \quad \forall i \in \I , k\in\S_i;
\notag\\
& & & ~~~   z_j \in \{0, 1\},\quad \forall j \in \J ;\notag\\
& &  & ~~~ y_k \in \{0, 1\},\quad \forall   k \in \D.\notag
\end{align}

\begin{theorem}\label{theorem:1}
Problem A1 and Problem A2 are equivalent.
\end{theorem}

This theorem can be proved by the above two lemmas directly.
Moreover, by transforming the Problem A1 into an equivalent and solvable Problem A2, we can adopt the classic algorithms for Knapsack problems directly into our problem.
For notational convenience, we denote the optimal solution of the Problem A2 (or A1) by $\{\x^o, \y^o, \z^o \}$.

\section{Information Asymmetry}
\label{sec:asymmetry}

In this section, we consider the information \emph{asymmetry} scenario, where the data requirement and valuation of a task are the private information of the task planner, and the sensing capability, sensing cost, and budget are the private information of each user, both of which cannot be observed by the platform.
Hence, the key problem is to design a \emph{truthful} mechanism to elicit the private information of task planners and users, and meanwhile to achieve the optimal task-data-user assignment as under information symmetry.

\subsection{Double Auction Framework}

Inspired by the VCG mechanism \cite{vcg}, we propose a VCG-based double auction mechanism for eliciting the private information of users and task planners.
In the proposed double auction, the MCS platform acts as the \emph{auctioneer},\footnote{The auctioneer can also be acted by any other third-party network node.} employing mobile users (\emph{bidders} on one side) for sensing different data items and selling the sensed data items to the required tasks (\emph{bidders} on the other side).
Different from a traditional VCG mechanism where the private information often resides on one side (either sellers or buyers) of the market, in our model the private information resides on both the user side (sellers) and the task plan side (buyers).

A typical VCG auction framework mainly consists of an \emph{assignment rule} (e.g., for deciding the assignment of buyers and sellers) and a \emph{payment rule} (e.g., for deciding the payments of buyers) \cite{vcg}. The payment rule is carefully designed such that bidders  will report the private information truthfully.
In our double auction framework, due to the two-sided private information, we need to design not only a \emph{payment rule} (for deciding the payments of task planners), but also a \emph{reward rule} (for deciding the rewards for users).
The payment rule is used to guarantee the truthful information disclosure of task planners, while the reward rule is used to
guarantee the truthful information disclosure of users.

Before presenting the detailed double auction rule, we first provide some important notations.
Denote
\begin{equation*}
\t_j \eq \{ \D_j, v_j \}  \mbox{~~~~and~~~~} \t_i \eq \{ \S_i, \c_i, C_i \}
\end{equation*}
as  \emph{true} information of task $j \in \J$ and user $i \in \I$. Denote
\begin{equation*}
\b_j \eq \{ \widetilde{\D}_j, \widetilde{v}_j \}  \mbox{~~~~and~~~~} \b_i \eq \{ \widetilde{\S}_i, \widetilde{\c}_i, \widetilde{C}_i \}
\end{equation*}
as the \emph{reported} information (bids) of task $j \in \J$ and user $i \in \I$, respectively.
For   convenience, we further denote $\bt \eq (\b_j ,j\in\J)$ and $\bu \eq (\b_i, i\in\I) $ as the bids of all task planners and users, respectively.
Obviously, if the proposed auction is truthful, we will have:  $\b_j  = \t_j $ and $\b_i = \t_i$.


With a little abuse of notations, we denote
\begin{equation}
\x (\cdot) \eq (\x _i(\cdot),i\in\I)   \mbox{~~~~and~~~~}  \z (\cdot) \eq (z_j(\cdot),j\in\J)
\end{equation}
as the \emph{assignment rule}, where $\x (\cdot)$ is the user scheduling rule and
$\z (\cdot)$ is the task completion rule.
We further denote
\begin{equation}
\p (\cdot) \eq (p_j(\cdot),j\in\J)
\end{equation}
as the \emph{payment rule} for task planners, where $p_j(\cdot)$ denotes the payment of each task planner $j \in \J$.
Similarly, we denote
\begin{equation}
\pv(\cdot) \eq (r_i(\cdot),i\in\I)
\end{equation}
as the \emph{reward rule} for users, where $r_i(\cdot)$ denotes the reward for each user $i \in \I$.
Based on the above, we can write such an auction mechanism as follows:
\begin{equation}
\mech \eq \{ \x (\cdot),\ \z (\cdot),\ \p (\cdot),\ \pv (\cdot) \}.
\end{equation}
Note that $\x (\cdot)$, $\z (\cdot)$, $\p (\cdot)$, and $\pv (\cdot)$
are all functions of   $\bu$ and $\bt$, hence can also be written as $\x (\bu, \bt)$, $\z (\bu, \bt)$, $\p (\bu, \bt)$, and $\pv (\bu, \bt)$.


\subsection{Truthful Double Auction}

Now we provide the detailed assignment rule, payment
rule, and reward rule for our proposed double auction $\mech$, based on the key idea of VCG auction~\cite{vcg}.

\begin{definition}[Assignment Rule]
The assignment rule $ \x (\bu, \bt)$ and  $\z (\bu, \bt)$ is given by:
 \begin{equation}\label{eq:1}
 \x (\bu, \bt) = \x ^o(\bu, \bt)
 \end{equation}
 and
 \begin{equation}\label{eq:2}
 \z   (\bu, \bt)= \z ^o(\bu, \bt),
 \end{equation}
 where $\{\x ^o(\bu, \bt), \z ^o(\bu, \bt)\}$ is the optimal solution to Problem A2 by replacing the true information $(\t_i, i\in \I )$ and  $ ( \t_j , j\in \J)$ with the reported bids $\bu $ and $ \bt$.
\end{definition}

\begin{definition}[Payment Rule]
The payment rule $\p (\bu, \bt)$  for task planners is given by:
 \begin{equation}\label{eq:3}
 \p (\bu, \bt) =  \left(p_j^o (\bu, \bt), j\in \J \right),
 \end{equation}
where the payment of task planner $j$ is
\begin{equation*}
 \begin{aligned}
p_j^o  (\bu, \bt) \eq & ~  W^o_{-j}(\bu, \bt)
\\
& - \sum\limits_{l \in \J/\{j\}} z_l^o (\bu, \bt) \cdot v_l
\\
& + \sum\limits_{i \in \I } \sum\limits_{k \subseteq \S_i}   x_{i,k}^o (\bu, \bt)  \cdot c_{i,k},
 \end{aligned}
\end{equation*}
and $ W^o_{-j}(\bu, \bt) $ is the maximum social welfare (defined on bids $\bu $ and $ \bt$) excluding task planer $j$.\footnote{\revv{Specifically, $W^o_{-j}(\bu, \bt) $ is the maximizer of Problem A2, by replacing the true information $(\t_i, i\in \I )$ and  $ ( \t_j , j\in \J)$ with the reported bids $\bu $ and $ \bt$, and meanwhile excluding task planer $j$.}}
\end{definition}

\begin{definition}[Reward Rule]
The reward rule $\pv (\bu, \bt)$  for users is given by:
 \begin{equation}\label{eq:4}
 \pv (\bu, \bt) =  \left(r_i^o (\bu, \bt), i\in \I \right),
 \end{equation}
where the reward to user $i $ is
\begin{equation*}
 \begin{aligned}
r_i^o (\bu, \bt) \eq &  \sum\limits_{j \in \J} z_j^o (\bu, \bt) \cdot v_j
\\
& - \sum\limits_{l \in \I /\{i\} } \sum\limits_{k \subseteq \S_l}   x_{l,k}^o (\bu, \bt)  \cdot c_{l,k}
\\
&
 - W^o_{-i}(\bu, \bt),
 \end{aligned}
\end{equation*}
and $ W^o_{-i}(\bu, \bt) $ is the maximum social welfare (defined on bids $\bu $ and $ \bt$) excluding user $i$.
\end{definition}

It is easy to prove that the double auction $\mech$ defined on Definition 1--3 is truthful and optimal (i.e., achieving the same optimal social welfare as under information symmetry).

\begin{theorem}\label{theorem:2}
The double auction mechanism $\mech$ given by \eqref{eq:1}--\eqref{eq:4} is truthful and optimal.
\end{theorem}

The proof for truthfulness is standard and can be referred to our technical report \cite{report}. The optimality can be easily shown by Definition 1, together with the truthfulness.

\subsection{Budget Balance}

Now we discuss the budget balance property of the proposed auction, which is important for incentivizing the auctioneer (platform) to organize such an auction.

Specifically, an auction is said to be \emph{(weakly) budget balance}, if the total payment collected from the task planners is no smaller than the total reward assigned to the users, hence the platform will not lose money by organizing such an auction.\footnote{An auction is said to be strictly budget balance, if the total payment collected from the task planners equals the total reward assigned to the users.}
Therefore, budget balance is a highly desirable property for our auction, otherwise the platform may lose the interest of organizing such an auction.

Unfortunately, our proposed double auction is \emph{not} budget
balance, mainly due to the data reuse among tasks.
Specifically, the data reuse among tasks reduces the competition among tasks, hence potentially reduces the payments of task planners (which is a common result in VCG), leading to the undesired budget imbalance.
This can be shown by the following simple example: (i) Two tasks requiring a same data item with $v_1 = 0.5$ and $v_2 = 0.6$, and (ii) One user can sense the data with cost $c = 0.2$. According to the assignment rule in \eqref{eq:1} and \eqref{eq:2}, the user will be scheduled the data item and both tasks will be completed, hence the maximum social welfare is $0.5+0.6-0.2=0.9$.
According to the reward rule in \eqref{eq:4}, the user will receive a reward of $ 0.5+0.6 - 0 - 0 = 1.1$.
According to the payment rule in \eqref{eq:3}, task 1 will be charged a payment of $0.4 - 0.6 +0.2 = 0$, and task 2 will be charged a payment of $0.3 - 0.5 +0.2 = 0$.
{This coincides with the common results in classic VCG mechanisms, where a task not generating harmful impact to the rest of the market often does not need to pay money.} 
Obviously, in this example, the platform losses a total money of $1.1$.

To this end, we introduce a \emph{reserve price} for each data item in the proposed double
auction, which denotes the minimal payment that a task planer has to pay for each data item.
Let $\pi_k \geq 0 $ denote the reserve price for each data item $k\in \D$.
Then, for each task planner $j \in \J$, his minimum payment (if task $j$ is completed) can be calculated by:
 \begin{equation}\label{eq:ppjj}
\underline{p}_j = \sum_{ k \in \D_j } \pi_k.
 \end{equation}

Based on the above, we propose the following new payment rule $\p^\dag (\bu, \bt)$ for task planners.
\begin{definition}[Payment Rule with Reserve Price]
The new payment rule $\p^\dag (\bu, \bt)$ for task planners is given by:
 \begin{equation}\label{eq:3a}
 \p^\dag (\bu, \bt) =  \left(p_j^\dag (\bu, \bt), j\in \J \right),
 \end{equation}
where the payment of task planner $j$ is
\begin{equation*}
p_j^\dag  (\bu, \bt) \eq \max\{ p_j^o  (\bu, \bt)  , \  \underline{p}_j\},
\end{equation*}
where $p_j^o  (\bu, \bt) $ is given in Definition 3.
\end{definition}

We will show that the double auction $\mech^\dag $ with the reserve price given in \eqref{eq:3a} is still truthful, but may be not optimal.

\begin{theorem}\label{theorem:3}
The double auction mechanism $\mech^\dag$ given by \eqref{eq:1}\eqref{eq:2}\eqref{eq:4}\eqref{eq:3a} is truthful (but not optimal).
\end{theorem}

The proof for truthfulness is still standard and can be referred to  our technical report \cite{report}.
The impact of the reserve price on optimality can be shown as follows.
With the reserve price, some task planners, i.e., those with a valuation lower than the minimum payment given in \eqref{eq:ppjj}, will decide to not join the auction.
Hence, the maximum  social welfare may be reduced.
Therefore, there is a tradeoff between the social efficiency and the budget balance.
A larger reserve price may lead to a better budget balance, but to a worse social efficiency.
We will show the impacts of reserve price on the budget balance and the social efficiency via simulations.


\section{Simulation Results}
\label{sec:simulation}

Now we provide simulation results to evaluate the performance of our proposed double auction mechanism.

\subsection{Simulation Setup}

To compare our proposed double auction with the randomized auction in \cite{Jiang2}, we consider a similar simulation setting as in \cite{Jiang2}, with $I=8$ tasks, $I=8$ users, and $K \in \{5,10,15,20\}$ target data items.
Each data item is location-based and distributed in an area of $10$km$\times10$km randomly and uniformly.
Each user is associated with a random location and can sense the data items within a distance of $5$km to his location.
Each task requires $5$ data items randomly picked from the whole target data set.
The sensing cost $c_{i,k}$ of each user $i$ for each data item $k$ is selected from $[0, 1]$ randomly and uniformly, and the total sensing budget $C_i$ of user $i$ is selected from $[0, 5]$ randomly and uniformly.
The valuation  $v_{j,k}$ of each task $j$ for each data item $k$ that it requires is selected from $[0,1.5]$ randomly and uniformly, and the total valuation $v_{j}$ of task $j$ is the sum of valuations on all required data items.


\begin{figure*}[t]
  \begin{minipage}[t]{.32\linewidth}
  \centering
  \includegraphics[scale=0.32]{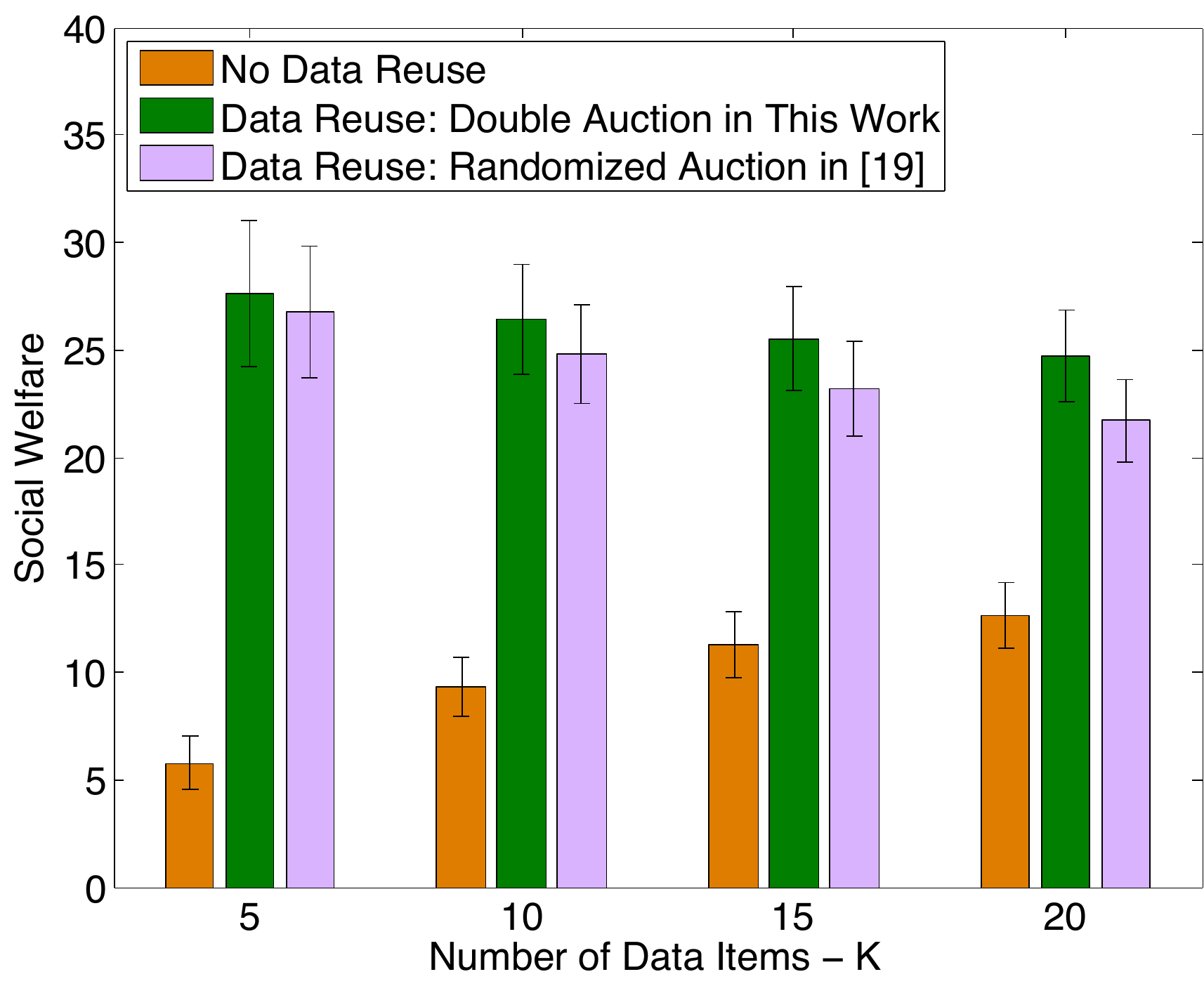}
   \caption{Social Welfare with/without Data Reuse}\label{Fig:xx1}
  \end{minipage}
  ~~~
  \begin{minipage}[t]{.32\linewidth}
  \centering
  \includegraphics[scale=0.32]{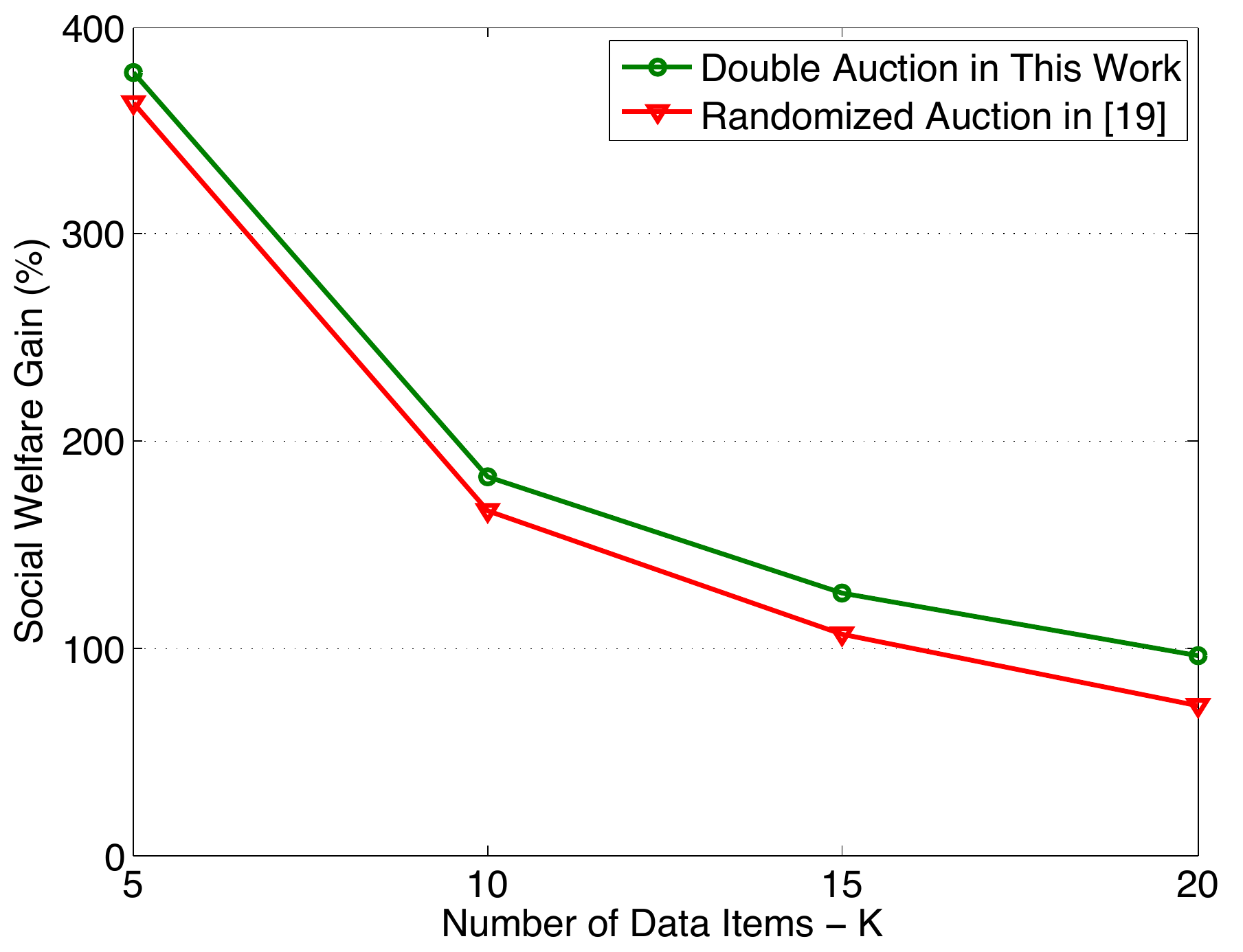}
   \caption{Social Welfare Gain due to Data Reuse}\label{Fig:xx2}
  \end{minipage}
  ~~~
  \begin{minipage}[t]{.32\linewidth}
  \centering
  \includegraphics[scale=0.32]{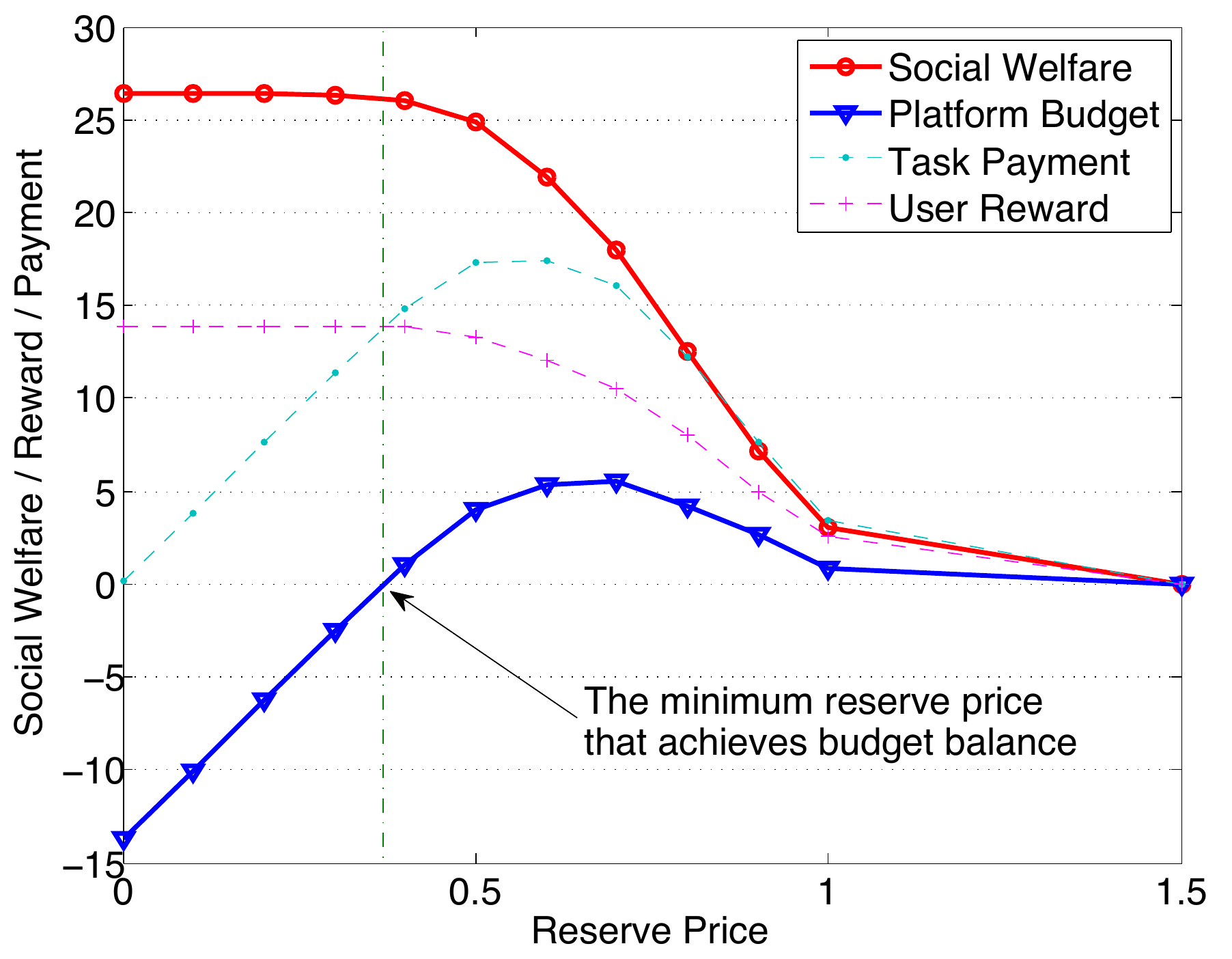}
   \caption{Platform Budget Balance}\label{Fig:xx3}
  \end{minipage}
\end{figure*}

\subsection{Social Welfare Gain}

We first illustrate the social welfare gain due to data reuse among tasks,  comparing with those without data reuse.

Figure \ref{Fig:xx1} shows the maximum social welfare without data reuse and with data reuse among tasks, under different number of data items.
For the case with data reuse, we further compare the social welfares achieved by our proposed double auction (which is provably optimal) and by the randomized auction in \cite{Jiang2} (which is approximately optimal).
We can see that the social welfare increases with the number of data items without data reuse, while decreases with the number of data items with data reuse.
The reason is as follows.
With a smaller set of data items, tasks are more likely to require the same data item.
Hence, with data reuse, they can potentially reuse a larger set of common data items, resulting in a lower sensing cost and hence a higher social welfare; without data reuse, however, the competition among tasks become more intensive (as the common data has to be sensed distinctly for each task), leading to a potentially lower social welfare.

Figure \ref{Fig:xx2} further shows the social welfare gain due to data reuse among tasks, comparing with those without data reuse.
We can see that the social welfare gain can be up to $380\% \sim 100\%$ by our proposed double auction, and $370\% \sim 80\%$ by the randomized auction in \cite{Jiang2}, when the number of data items changes from $5$ to $20$.
That is, our proposed double auction has an average of
$10\%$ performance gain over the randomized auction in \cite{Jiang2}.
When the number of data items is very large, there will be no common data requirement among tasks (hence no data reuse), and thus the social welfare gain due to data reuse will go to zero, which implies that the social welfares with and without data reuse are same.

\subsection{Platform Budget}

We now illustrate the impacts of the reserve price on the platform budget and  the social welfare.
We will show that a well-designed reserve price can achieve a desired tradeoff between   budget balance and social efficiency.~~~~~~~~~~~~~~~~~~~~~~

Figure \ref{Fig:xx3} shows the social welfare and the platform budget under different reserve prices.
For clarity, we also present the total task payment and the total user reward in the figure, and the platform budget is just the difference between the total task payment and user reward.
We can see that the social welfare always decreases with the reserve price,
while the platform budget first increases and then decreases with the reserve price.
The reason is as follows.
Note that a task planner will leave the auction, if his task valuation is lower than the minimum payment defined in \eqref{eq:ppjj}, i.e., the
sum of reserve price on all required data items.
Thus, with the increasing of the reserve price, more task planners are likely to leave the auction, resulting in a lower social welfare.
Moreover, when the reserve price increases from a small level, most task planners still stay in the auction, and hence the platform budget increases due to the increased payment from task planners;
when the reserve price increases from a high level,
many task planners leave the auction, hence the platform budget decreases due to the decreased number of task planners.
Besides, when the reserve price is very large (e.g., $1.5$ in the figure), almost all of the task planners will leave the auction, leading to a zero social welfare and a zero platform budget.

Figure \ref{Fig:xx3}  also shows the minimum reserve price that achieves the platform budget balance, i.e., that leading to a zero budget for the platform (i.e., $0.38$ in the figure).
We can see that under such a reserve price (which achieves the strict budget balance), the social welfare loss is less than $2\%$, comparing with the maximum social welfare under the  zero reserve price.

%


\section{Conclusion}
\label{sec:conclusion}

In this work, we consider a novel MCS framework, where different tasks may have the common data requirement and can reuse the common data through a MCS platform. 
We study the optimal assignment among mobile users and sensing tasks (with data sharing) under both information symmetry and asymmetry. 
In particular, we propose a truthful and optimal double auction mechanism under information asymmetry. 
We further introduce a reserve price to achieve a desired tradeoff between the budget balance and the social efficiency. 
There are several interesting directions for future research. First, it is meaningful to consider more practical valuation model for tasks and cost model for users.
Second, it is also important to study the approximate algorithm for solving the inherit NP-hard assignment problem.

\newpage

\newpage

Online technical report for ``\textbf{\emph{A Double Auction Mechanism for Mobile Crowd
Sensing with Data Reuse}}'' submitted to  {IEEE GLOBECOM 2017}.
Outline of the report:

{\it
\begin{itemize}
  \item \ref{app:1}: Proof for Lemma \ref{lemma:1}
  \item \ref{app:2}: Proof for Lemma \ref{lemma:2}
  \item \ref{app:3}: Proof for Theorem \ref{theorem:1}
  \item \ref{app:4}: Proof for Theorem \ref{theorem:2}
  \item \ref{app:5}: Proof for Theorem \ref{theorem:3}
\end{itemize}
}

\appendix

\section{Appendix}

\subsection{Proof for Lemma \ref{lemma:1}}\label{app:1}

\begin{proof}
We consider two cases: (i) $x_{i,k} = 0, \forall i \in \I_k$, and (ii) there exists at least one $i\in \I_k$ with $x_{i,k} = 1$.
In the first case, we have $ y_k = 0$ by both \eqref{eq:ykkk} and \eqref{eq:ykkk-eq}.
In the second case, we have $ y_k = 1$ by both \eqref{eq:ykkk} and \eqref{eq:ykkk-eq}.
Hence, \eqref{eq:ykkk} and \eqref{eq:ykkk-eq} are equivalent with each other.
\end{proof}

\subsection{Proof for Lemma \ref{lemma:2}}\label{app:2}

\begin{proof}
We consider two cases: (i) $y_k = 1, \forall k \in \D_j $, and (ii) there exists at least one $k \in \D_j $ with $y_k = 1$.
In the first case, we have $  z_j = 1$ by \eqref{eq:zj}. It is easy to check that the optimal $z_j$ under the constraint \eqref{eq:zj-eq} is also $  z_j = 1$ due to the definition of $V(\z)$ in \eqref{eq:Vz}.
In the second case, we have $ z_j =  0 $ by both \eqref{eq:zj} and \eqref{eq:zj-eq}.
Hence, relaxing \eqref{eq:zj} to \eqref{eq:zj-eq} does not affect the optimal solution.
\end{proof}

\subsection{Proof for Theorem  \ref{theorem:1}}\label{app:3}

\begin{proof}
By Lemma \ref{lemma:1} and Lemma \ref{lemma:2}, we can prove the theorem directly.
\end{proof}

\subsection{Proof for Theorem  \ref{theorem:2}}\label{app:4}

\begin{proof}
\emph{Proof for Truthfulness:} It is easy to see that the payment rules given by \eqref{eq:3} and \eqref{eq:4} follow the basic principle of VCG mechanism \cite{vcg}, where each bidder (task planner or user) is charged by his critical bid, i.e., the harm they cause to other bidders. According to the classic results in \cite{vcg}, each bidder will report his private information truthfully.

\emph{Proof for Optimality:} Based on the above, each bidder will report his private information truthfully. Thus, the assignment rule given by \eqref{eq:1} and \eqref{eq:2} is optimal, that is, it maximizes the social welfare.
\end{proof}

\subsection{Proof for Theorem  \ref{theorem:3}}\label{app:5}

 \begin{proof} Similar as the proof for Theorem  \ref{theorem:2}, we can easily find that the payment rules given by \eqref{eq:4} and \eqref{eq:3a} follow the basic principle of VCG mechanism \cite{vcg}, where each bidder (task planner or user) is charged by his critical bid, i.e., the harm they cause to other bidders. Thus, by the classic results in \cite{vcg}, each bidder will report his private information truthfully.
\end{proof}

\end{document}